# Quantitative mapping of chemical defects at charged grain boundaries in a ferroelectric oxide

K. A. Hunnestad[1], J. Schultheiß[1], A. C. Mathisen[2], I. Ushakov[1], C. Hatzoglou[1], A. T. J. van Helvoort[2], and D. Meier[1]

[1] Department of Materials Science and Engineering, Norwegian University of Science and Technology (NTNU), 7491 Trondheim, Norway

[2] Department of Physics, Norwegian University of Science and Technology (NTNU), 7491 Trondheim, Norway

**Polar discontinuities and structural changes at oxide interfaces can give rise to a large variety of electronic and ionic phenomena. Related effects have been intensively studied in epitaxial systems, including ferroelectric domain walls and interfaces in superlattices. Here, we investigate the relation between polar discontinuities and the local chemistry at grain boundaries in polycrystalline ferroelectric $ErMnO_3$. Using orientation mapping and different scanning probe microscopy techniques, we demonstrate that the polycrystalline material develops charged grain boundaries with enhanced electronic conductance. By performing atom probe tomography measurements, we find an enrichment of erbium and a depletion of oxygen at all grain boundaries. The observed compositional changes translate into a charge that exceeds possible polarization-driven effects, demonstrating that structural phenomena rather than electrostatics determine the local chemical composition and related changes in the electronic transport behavior. The study shows that the charged grain boundaries behave distinctly different from charged domain walls, giving additional opportunities for property engineering at polar oxide interfaces.**

Grain boundaries in polycrystalline materials represent quasi-2D systems that naturally form during processing, separating adjacent regions of different crystallographic orientation. Due to the discontinuous structural changes across the boundaries, they can display a completely different configuration and symmetry than the surrounding grains. In analogy to domain walls and artificially designed interfaces, the grain-boundary-related structural differences give rise to distinct properties and functionalities, opening an avenue for engineering electronic responses and physical phenomena at the local scale. For example, grain boundaries in polycrystalline oxides, such as $CeO_2$[1] and $SrTiO_3$[2], display a substantially different electronic and ionic transport behavior than the bulk[3], and grain boundaries in $LaAlO_3$ were found to be flexoelectric[4]. Furthermore, grain boundaries play an important role for applications as reflected by ZnO polycrystals, which are used as varistors[5]. In many cases, the newly formed properties are attributed to, or go hand in hand with, an accumulation (or depletion) of point defects[1,6,7]. Facilitated by advances in transmission electron microscopy (TEM), the structure and elemental composition at grain boundaries has been studied[8]. TEM measurements allowed, for instance, for tracing the ten times increased grain boundary conductivity in polycrystalline $Gd_{0.2}Ce_{0.8}O_2$ back to a local variation in the Ce and Gd concentration by more than 10 at. %[1]. In $(Li_{1/3x}La_{2/3-x})TiO_3$, it was found that only grain boundaries of a certain structural configuration display reduced Li-ion conductivity, which was explained by the presence of oxygen vacancy accumulations[6]. Despite the remarkable success, TEM-based studies are limited to defect concentrations in the order of a few at. %, and the investigation of 2D projections along certain zone axes. To enable a complete three-dimensional (3D) characterization of grain boundaries and expand measurements towards lower defect concentrations, different experimental strategies are required.

A particularly powerful imaging technique that facilitates 3D measurements with high spatial resolution is atom probe tomography (APT). Combining high chemical sensitivity and accuracy with atomic-scale spatial resolution, APT allows for mapping and quantifying defect concentrations with unprecedent accuracy in 3D and can support TEM-based chemical analysis. APT is commonly applied in metallurgy to gain insight into the local chemical composition at grain boundaries, and it has been used to investigate the chemistry of precipitates, dislocations, and other nanostructures[9,10]. In contrast to the extensive work in the field of metallurgy, grain boundaries in oxide materials have been studied much less by APT[7,11]. One intriguing example is high-resolution measurements on grain boundaries in lightly doped ceria oxide, which revealed a space charge region caused by an accumulation of trace impurities with a concentration of less than 1 at. %[7]. This pioneering work demonstrated the general potential of APT for studying grain boundaries in oxides and motivates our work, which focusses on charged grain boundaries that arise in polar oxides. Going beyond the so far investigated systems, polycrystalline polar materials, such as ferroelectrics and pyroelectrics, form a spontaneous electric polarization. This polarization can lead to interfaces with bound charges and promote the migration of charged point defects with so far unexplored consequences for the grain boundaries.

Here, we characterize the electronic properties and quantify the defect chemistry at charged grain boundaries in ferroelectric ErMnO$_3$ polycrystals. Kelvin probe and conductive atomic force microscopy studies show that the electrostatic potential and electronic conductance at the grain boundaries is different from the surrounding grains. By performing APT measurements, we reveal an Er enrichment and O depletion at the grain boundaries, with subtle compositional changes of about 1-2 at. % compared to the interior of the grains. This deviation indicates a change in the cation-anion ratio at the grain boundary, which locally violates charge neutrality. The change is too large to be explained based on electrostatic effects alone as measurements at different grain boundaries show.

**Results and Discussion**.

**Micro- and nanoscale characterization.** ErMnO$_3$ belongs to the family of hexagonal manganites ($R$MnO$_3$, $R$=Sc, Y, and Ho-Lu). The material is a uniaxial ferroelectric with a spontaneous polarization, $P \approx 5.6$ µC/cm², oriented parallel to the crystallographic $c$-axis of the hexagonal unit cell[12]. The polarization emerges as a symmetry enforced by-product of a structurally driven phase transition at 1429 K[13], resulting in a variety of interesting physical features, including topologically protected vortex cores[14,15], stable charged ferroelectric domain walls[16,17], and unusual domain scaling behavior[18,19]. While the majority of the research focused on single crystals, polycrystalline $R$MnO$_3$ as studied in this work (see methods for processing details) are much less explored, despite their intriguing properties. Previous studies on polycrystalline $R$MnO$_3$ thin films, for example, showed that grain boundaries represent an additional degree of freedom that can be utilized to realize resistive switching[20] or nonvolatile memory applications[12]. More recently, bulk polycrystalline ErMnO$_3$ was found to exhibit an inverted domain-size/grain-size scaling behavior, enabled by the strain-driving unfolding of topologically protected vortex structures and the annihilation of vortex cores at grain boundaries[21]. The electronic properties and chemistry of the grain boundaries in polycrystalline hexagonal manganites, however, remain unknown and additional insight is highly desirable to understand their impact on the material's electric and dielectric responses, as well as the mechanical behavior.

To characterize the electronic properties of the grain boundaries in our polycrystalline ErMnO$_3$ sample, we combine different scanning probe microscopy (SPM) techniques, measuring the local piezoresponse, electrostatic potential and electronic conductance as summarized in Figure 1. Figure 1A displays a representative piezoresponse force microscopy (PFM) image, which is recorded with alternating-current (ac) voltage of 5 V (peak-to-peak) applied to the back of the sample with a frequency of 314.7 kHz. The PFM image displays the out-of-plane piezoelectric response, revealing the characteristic ferroelectric domain structure of polycrystalline ErMnO$_3$ as discussed elsewhere[21]. Consistent with literature, the vortex-like domain structure in the grains is mostly suppressed (for further details on the in Figure 1A, using an ac voltage of 3 V at 66.7 kHz. A pronounced KPFM contrast is observed (dark = low potential; bright = high potential), indicating substantial variations in the surface potential[7,22,23]. In particular, we find that the surface potential is enhanced at the grain boundaries, which is also shown by the line profile in Figure 1C, indicating an about 15 mV higher potential at the grain boundary relative to the surrounding grains. We note that the corresponding topography profile (Figure 1C) is flat on the nanometer scale without any detectable structural anomaly, which leads us to the conclusion that the KPFM contrast originates from a variation in the local electric properties rather than topographic features. This conclusion is corroborated by conductive atomic force microscopy (cAFM) measurements at 2 V as presented in Figure 1D; although no substantial variation in surface topography is detected, the cAFM data shows enhanced electronic conductance at the position of the grain boundaries, reflecting locally different electric properties.

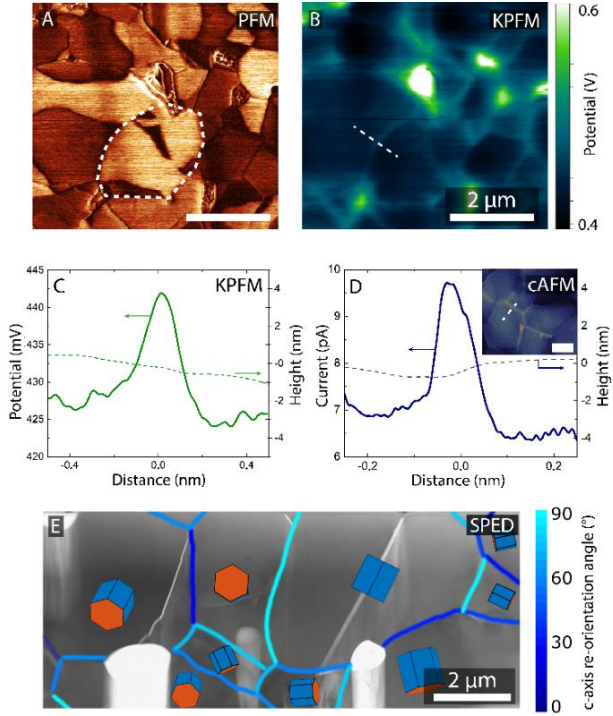

**Figure 1: Microstructural characterization.** (A) Out-of-plane PFM image, displaying pronounced bright and dark contrast consistent with grain-to-grain variations and ferroelectric 180° domains as discussed in ref. 21. The dashed line highlights the boundary of a representative grain. (B) KPFM map of the same area as shown in (A), where grain boundaries display a brighter contrast (higher potential) than the bulk. A profile across one grain boundary is extracted along the dashed line and shown in (C), along with the topography. (D) cAFM profile and map (inset) of a grain boundary along with the topography. Scale bar is 500 nm. (E) SPED is used to obtain the crystal orientation of the grains, as represented by a hexagonal prism with the ferroelectric polarization lying along the hexagonal $c$-axis. The calculated $c$-axis reorientation angle between neighboring grains is indicated by different shades of blue.

domain structure, the reader is referred to ref. 21). Based on our PFM analysis and simultaneously recorded topography data (not shown), we measure an average grain size of about 1.5±0.8 μm (for reference, the boundary of one grain is marked by the dashed line in Figure 1A). Complementary Kelvin probe force microscopy (KPFM) data is presented in Figure 1B; the scan is taken in non-contact mode at the same location as the PFM image

To investigate the relation between the unusual electronic properties and emergent polar discontinuities at the grain boundaries due to the uniaxial ferroelectricity in ErMnO$_3$ ($P \parallel c$), we analyzed grain orientations by scanning precession electron diffraction (SPED) combined with template matching[24] (Figure 1E). In Figure 1E, we illustrate the orientation of different grains by a sketch of the hexagonal prism and indicate the calculated 3D angular difference in $c$-axis across the different grain boundaries. High-resolution TEM data (Figure S1) shows that the grain boundary extends at most over a few unit cells without an apparent intergranular boundary phase present. The ferroelectric 180° domain walls[16]. From the SPED data alone, however, it is not possible to say whether grain boundary bound charges are sufficient to explain their distinct electronic response.

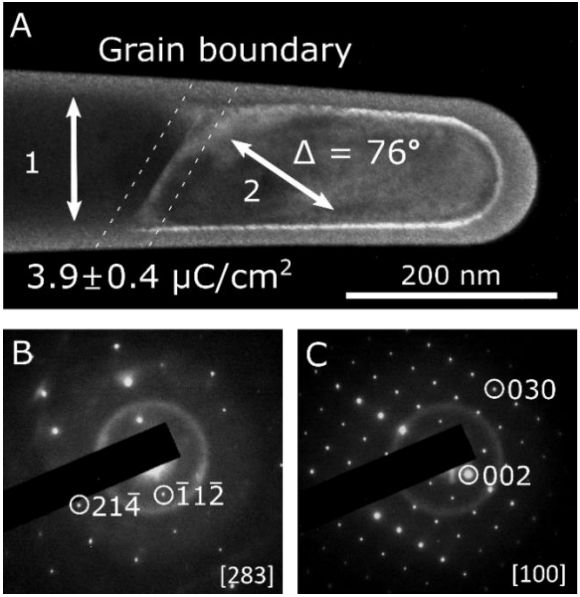

**Figure 2: Grain boundary extraction and analysis.** (A) DF-TEM of an APT needle showing to the grain boundary configuration. The crystal orientation of the two grains is identified via the electron diffraction images in (B) and (C), with the viewing direction oriented to keep grain 2 on the (100) zone axis and the 2D projected polarization directions drawn in with white arrows. The resulting change in *c*-axis across the grain boundary is about 76°, corresponding to a bound charge $|\rho|$ = 3.9 ± 0.4 µC/cm². For the DF-TEM image in (A), the 002 reflection in (C) was used.

**APT analysis of charged grain boundaries.** To understand the origin of the electronic properties at the grain boundaries, we next analyze their chemical composition utilizing ATP. For this purpose, we extract individual grain boundaries in needle-shaped specimens with a tip radius of less than 100 nm from polycrystalline samples using a focused ion beam (FIB). The characteristic needle shape is required to produce the high electric field needed for field evaporation during the APT investigation. Dark-field (DF) TEM data of a representative APT needle with a single grain boundary is displayed in Figure 2A. The grain boundary is visible as an abrupt change in contrast between grain 1 and 2, lying between the two dashed lines in Figure 2A. The orientation of the two grains and their polar axes (white double-headed arrows in Figure 2A) are identified by acquiring selected area electron diffraction (SAD) patterns of each grain, displayed in Figure 2B and C. From the two manually indexed SAD patterns, we identify the angle across the grain boundary separating the *c*-axes of the hexagonal grains ($P \parallel c$), which is about 76° (error in the range of 5°-10°). Thus, we can conclude that the grain boundary carries a finite bound charge $\rho$, which arises from the discontinuity of the polarization component perpendicular to it: $\rho = (\boldsymbol{P_1} - \boldsymbol{P_2}) * \boldsymbol{n}$, where $\boldsymbol{P_{1,2}}$ is the polarization in the two grains, and $\boldsymbol{n}$ is the grain boundary normal unit vector[25]. Although TEM only gives a 2D projection of the grain boundary normal vector, and we do not resolve decoupling of the ferroelectric domains in adjacent grains, together with the random orientation of the grains, results in a complex charge state at the grain boundaries, defined by the local polarization direction of neighboring grains (Figure 1E). In general, for randomly oriented grains, the majority of grain boundaries is charged and, hence, requires charge compensation which is consistent with their unusual electronic properties. Assuming an abrupt change in the polarization angle from an otherwise continuous phase, analogous to ferroelectric domain walls, bound charges with an upper limit of 11.2 µC/cm² can arise. The value corresponds to a fully charged grain boundary in head-to-head (→←) or tail-to-tail (←→) configuration, analogous to fully charged the sign of *P*, we can extract an estimate for the local charge state, which is $|\rho| = 3.9 \pm 0.4$ μC/cm$^2$. We note that the TEM data in Figure 2A also reveals a thin amorphous layer on the surface of the needle with a thickness of about 20-30 nm, which originates from the FIB sample preparation[26] and is outside the reconstructed APT volume.

By combining a strong electric field (a few tens of V/nm) with femtosecond laser pulses (wavelength $\lambda$ = 355 nm), we field-evaporate the apex of the needle. Collecting the individual atoms enables us to reconstruct a full 3D atomic distribution map, which is displayed in Figure 3A. As highlighted in Figure 3B, each atom is identified as a specific ionic species via its mass-to-charge ratio, obtained by measuring the time-of-flight during the evaporation (see Figure S2). Considering all collected atoms in the dataset, the averaged chemical composition is obtained and quantified as 21.914 ± 0.005 at. % Er, 22.455 ± 0.005 at. % Mn, and 55.588 ± 0.004 at. % O, consistent with the nominal composition of our polycrystalline ErMnO$_3$ material, i.e., 20 at. % Er, 20 at. % Mn, and 60 at. % O. Note that the measured oxygen concentration is lower than the nominal value. It is well-known for APT that some elements have a lower detection efficiency due to molecular dissociation, generating neutral species or preferential evaporation[27]. Interestingly, a change in composition is detected in the lower part of the needle, which coincides with the location of the grain boundary as measured by TEM (Figure S3A). A zoom-in to the region of interest is presented in Figure 3C, where the grain boundary is visualized through an isosurface, highlighting regions with increased Er atomic density. Based on the correlated TEM and APT data, we conclude that the 2D features seen in the two data sets correspond to the same grain boundary, which is analyzed further in Figure 4A.

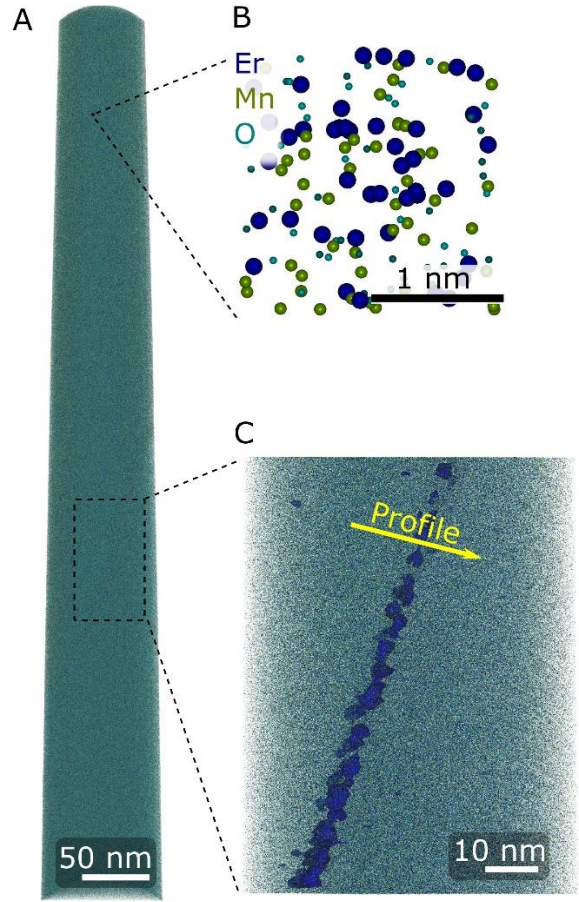

**Figure 3: 3D reconstructed APT data.** (A) Representative 3D reconstructed data from APT specimen after field evaporation. (B) A 2x2x2 nm$^3$ volume from the core of the needle, showing the positions of individual atoms and their color scheme. (C) An enlarged image of the region containing the grain boundary. The area where the compositional profile is obtained from is highlighted by the dashed lines. The grain boundary can be seen as a subtle contrast change in (A), whereas in (C), it is visualized through an Er density isosurface indicated in blue. The isosurface contains regions with more than 10 Er atoms /nm$^3$, up from an average of 6 Er atoms /nm$^3$.

**Chemistry of charged grain boundaries**. To gain quantitative information about the chemical composition, the compositional profile across the grain boundary (indicated by the arrow in Figure 3C) is shown in Figure 4A. The data indicates a chemical grain-to-grain variation, which might be a consequence of the diffusion-driven solid-state synthesis of the ErMnO$_3$ particles from Er and Mn oxides (see methods).[28] Most important for this work, we observe an additional change that is specific to the grain boundaries and independent of the grain-to-grain variation. Following the profile in Figure 4A from left to right, the oxygen concentration substantially drops at the position of the grain boundary. The drop below the level of either grain (shaded regions) indicates a depletion in oxygen at the grain boundary core of about 1 at. %. For the Er profile, the opposite trend is seen with an enrichment of about 1 at. % at the grain boundary core. In contrast to O and Er, only minor changes in the Mn concentration are observed.

To evaluate whether the observed variations in chemical composition are specific to the analyzed grain-boundary or a general phenomenon, we repeat the APT analysis on several grain boundaries in needles extracted from different regions <100 μm apart from each other on the same ErMnO$_3$ polycrystal. Representative chemical composition profiles along two additional grain boundaries are presented in Figure 4B and C (corresponding TEM data is shown in Figure S3A-C and Figure S3D-F, respectively). For all the analyzed boundaries, we consistently observe an O depletion and Er enrichment in the order of 1 to 2 at. %. This observation leads us to the conclusion that the local change in chemical composition is common for the analyzed grain boundaries in our polycrystalline ErMnO$_3$.

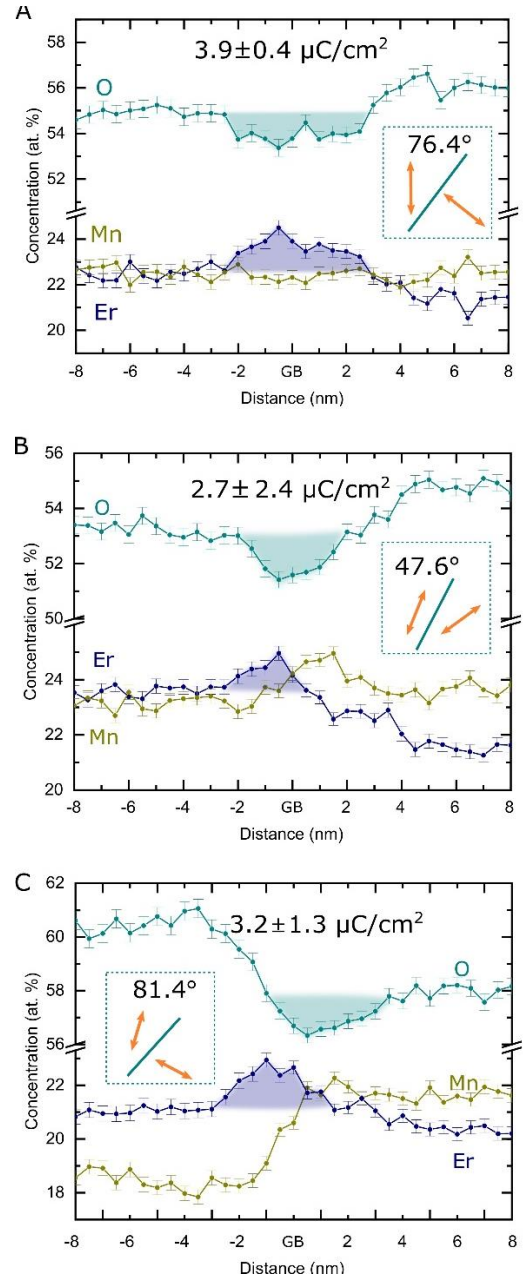

**Figure 4: Chemical composition across grain boundaries in ErMnO$_3$.** (A-C) The chemical composition of three grain boundaries in polycrystalline ErMnO$_3$, displaying the concentrations of the elements Er, Mn, and O across the grain boundary. The main difference between grains and grain boundaries is a consistent depletion in O and enrichment in Er of about 1 at. %, respectively, as highlighted by the shaded regions. The angle between *c*-axes obtained from TEM (Figure S3A-C correspond to (B) and Figure S3D-F correspond to (C)) is given in the boxes inside each panel, and the relation between polarization axes and grain boundary is sketched. The bound charge is calculated for each grain boundary, displayed at the top with a lower and upper limit.

Thus, the APT analysis reveals a propensity of Er to accumulate at the grain boundaries, which is not observed for Mn, i.e., the species with the smaller ionic radius ($r_{Er}$ = 0.945 Å, $r_{Mn}$ = 0.58 Å[29]). Interestingly, the simultaneous enhancement in Er together with a depletion in O violates charge neutrality. To compensate the positively charged $Er^{3+}$ atoms, negatively charged $O^{2-}$ interstitials would need to accumulate at the grain boundary, which is opposite to what we measure by APT. The observed change in the cation-anion ratio at the grain boundary implies a positively charged boundary, which is consistent with an enhanced potential measured by KPFM (Figure 1B and C). Under the simple assumption that every missing oxygen atom gives a charge of $2e^-$, the measured oxygen depletion is equivalent to a charge >130 µC/cm$^2$ (see methods for calculation details). This value is more than an order of magnitude larger than the maximum bound charge that can arise at grain boundaries due to the spontaneous polarization of ErMnO$_3$, which clearly discards electrostatics as the driving mechanism. This leads us to the conclusion that other effects, such as structural or chemical grain-to-grain variations, dominate the ionic redistribution observed at the grain boundaries in the ErMnO$_3$ polycrystal – and, hence, their electronic response – which is distinctly different from charged domain walls. Although the resulting compositional change at the grain boundary core is small, our SPM measurements (Figure 1) show that it has drastic consequences for the local electrostatics and transport properties, impacting the functionality of grain boundaries in polycrystalline ErMnO$_3$.

## Conclusions

In summary, we showed that charged grain boundary with enhanced electrical conductance arise in ErMnO$_3$ polycrystals and quantified the grain-boundary-related compositional fluctuations and impurity accumulations. We consistently find a change of the cation-anion ratio at the grain boundary core where the Er is enriched by about 1-2 at.% and O is equally depleted. In turn, this violates charge neutrality which cannot be explained based on electrostatics, indicating the dominant role of local structural effects and chemical grain-to-grain variations. The behavior is distinctly different from charged ferroelectric domain walls, where the polarization discontinuity determines the electronic properties. Our results give new insight into the electronic properties and chemical composition of charged grain boundaries, showing additional opportunities for the property engineering of polar quasi-2D systems in ferroelectric oxides and nonmetallic inorganics in general.

## Methods

**Material Synthesis.** Synthesis of ErMnO$_3$ was done by a solid-state reaction utilizing dried Er$_2$O$_3$ (99.9%, Alfa Aesar, Haverhill, MA, USA, Lot No.: M23D038) and Mn$_2$O$_3$ (99.0%, Alfa Aesar, Haverhill, MA, USA, Lot No.: 61700342) as starting materials, followed by subsequent stepwise heat treatment (1270K-1370 K) for 12 hrs and sintering to achieve a dense material (1623 K for 4 hrs). Further details on the processing conditions can be found in ref. 30. The crystal structure and phase purity of our sample is confirmed by powder X-Ray diffraction (XRD) (Figure S4) indicating a pure hexagonal crystal structure with a space group symmetry *P*6$_3$*cm*, similar to single crystalline ErMnO$_3$[31].

**SEM/FIB.** A Helios NanoLab DualBeam FIB (Thermo Fisher Scientific, Massachusetts, USA) was used for preparing APT specimens and to image the bulk specimens with an acceleration voltage for the electron beam of 3 kV while recording the secondary electron emission. See ref. 32 for a detailed description of SEM imaging techniques of ferroelectric materials. The extraction procedure from bulk specimen to APT needle is described in ref. 33. Final APT specimens were placed on TEM grids to support correlative APT-TEM and low-voltage polishing (<5 kV) with a $Ga^+$ beam to minimize surface damage on the specimens.

**AFM.** For local electromechanical characterization, the sample was lapped with a 9 μm-grained (Logitech Ltd, Glasgow, UK) water suspension and polished using a silica slurry (Ultra-Sol®, Eminess Technologies, Scottsdale, AZ, USA). PFM and KPFM data were acquired with Cypher (Asylum Research, Oxford Instruments, USA), using an ASYELEC.01-R2 Ti/Ir coated Si tip (Oxford Instruments, USA), with a measured spring constant and resonance frequency of 1.58 N/m and 66.7 kHz, respectively. The PFM was performed in resonance mode, with contact resonance frequencies found to be 314.7 kHz and 689.4 kHz for vertical and lateral PFM signals, respectively. The cantilever deflection force setpoint was kept at around 50 nN during scanning. The sample back electrode was grounded, and an alternating voltage of 5 V was applied. The amplitude and phase of the cantilever movements were then measured. The KPFM was performed in a two-pass mode. The first pass mapped the topography using oscillating mode, with free air amplitude of about 265 nm, a setpoint amplitude of about 125 nm, and an oscillating frequency of 66.7 kHz. On the second pass, the tip was lowered by 70 nm (about 80 nm tip-surface distance), following the topography path while scanning. An alternating voltage of 3 V at 66.7 kHz was applied to the tip with grounded back electrode. Simultaneously, a direct voltage was applied in a such manner that cantilever oscillations of frequency 66.7 kHz was suppressed. To acquire the local conductance, cAFM was performed in contact mode with a voltage of 2 V applied to the back electrode.

**TEM.** Using a JEOL 2100F Field Emission Gun (FEG) microscope (JEOL, Tokyo, Japan), operated at 200 kV, selected specimens were investigated to check for the presence of grain boundaries. Electron diffraction patterns were acquired using a 2K Gatan multiscan CCD while selecting a ≈100 nm circular region corresponding to each grain. For imaging (DF-TEM) and HRTEM the same CCD was used, The SPED data was recorded using a 1° precession angle and an ASTAR set-up (Nanomegas, Brussels, Belgium) and Medipix direct electron detector (R1, QuantumDetectors, Oxford, UK). The SPED data stack was analyzed using the open-source package pyXem[34].

**APT.** APT data was acquired using a Cameca LEAP 5000XS (CAMECA, Gennevilliers, France), operated in laser pulsing mode. The laser pulses temporarily heat up the specimens, which combined with a direct-current (dc) electric bias field evaporates the ions. Spatial positions were recorded with a 2D detector while measuring the time-of-flight provides information on the mass of the ions. Laser pulses with a frequency of 250-500 kHz and energy of 30 pJ were used. Detection rate was set to 0.5-1.0 %, meaning that 5-10 atoms are recorded on average per 1000 pulses. Specimens were cooled down to 25 K during the whole analysis. For the reconstruction of raw APT data into 3D datasets, the software Cameca IVAS, version 3.6.12, was used with a voltage-based radial evolution. The parameters were optimized to ensure the interfaces appear flat and not curved. The peak at 16 Da in the mass spectrum (see Figure S2) could correspond to either $O^+$ or $O_2^{2+}$. Following the discussion in refs. 35 and 36 the peak is ranged as $O^+$ and not $O_2^{2+}$. A lower limit for the bound charge at the grain boundary was found by considering only the chemical change of the oxygen vacancy region: $\sigma = \rho * V_O * 2e^-$, where $\sigma$ is the bound charge, $\rho$ is the atomic density in $ErMnO_3$ (81.1 atoms/nm$^2$) over a 1 nm interface, $V_O$ is the integrated oxygen depletion obtained from APT and $e^-$ is the elementary charge.

ASSOCIATED CONTENT

**Supporting Information.**

High-resolution TEM data; Mass spectrum of a bulk sample; TEM data from additional grain boundaries; Si impurity observation; Si concentration plots from additional grain boundaries; XRD of bulk samples.

AUTHOR INFORMATION

**Corresponding Author**

Jan Schultheiß. E-mail: jan.schultheiss@ntnu.no
Dennis Meier. E-mail: dennis.meier@ntnu.no

**Author Contributions**

APT specimen preparation, data collection and analysis were performed by K.A.H., supervised by D.M. and A.T.J.v.H. Polycrystalline samples were synthesized and characterized by XRD by J.S.. TEM data was acquired by K.A.H., and diffraction data was collected and analyzed by A.C.M. and A.T.J.v.H.. SPM data were collected by I. U.. The project was devised by D.M. and coordinated by D.M. and J.S. The manuscript was written through contributions of all authors, led by K.A.H, J.S., and D.M. All authors have given approval to the final version of the manuscript.

ACKNOWLEDGMENT

The Research Council of Norway (RCN) is acknowledged for the support to the Norwegian Micro- and Nano-Fabrication Facility, NorFab, project number 295864, the Norwegian Laboratory for Mineral and Materials Characterization, MiMaC, project number 269842/F50, and the Norwegian Center for Transmission Electron Microscopy, NORTEM (197405/F50). Emil Christiansen is thanked for the support in collecting SPED data. K.A.H. and D.M. thank the Department of Materials Science and Engineering at NTNU for direct financial support. J.S. thanks the Alexander-von-Humboldt Foundation for support through a Feodor-Lynen research fellowship. D.M. acknowledges funding from the European Research Council (ERC) under the European Union's Horizon 2020 research and innovation program (Grant Agreement No. 863691) and further thanks NTNU for support through the Onsager Fellowship Program and NTNU Stjerneprogrammet. Hanne-Sofie Søreide is thanked for her support to the APT lab facilities, Sverre Magnus Selbach is acknowledged for helpful discussions.

ABBREVIATIONS

KPFM, Kelvin-probe force microscopy; cAFM, conductive atomic force microscopy; APT, atom probe tomography; TEM, transmission electron microscopy; FIB, focused ion beam, DF, dark-field; PFM, piezo-force microscopy; SAD, selected area electron diffraction; SPED, scanning precession electron diffraction; SPM, scanning probe microscopy.

# Quantitative mapping of chemical defects at charged grain boundaries in a ferroelectric oxide

K. A. Hunnestad[1], J. Schultheiß[1], A. C. Mathisen[2], I. Ushakov[1], C. Hatzoglou[1], A. T. J. van Helvoort[2], and D. Meier[1]

[1] Department of Materials Science and Engineering, Norwegian University of Science and Technology (NTNU), 7491 Trondheim, Norway

[2] Department of Physics, Norwegian University of Science and Technology (NTNU), 7491 Trondheim, Norway

**Supporting figures**

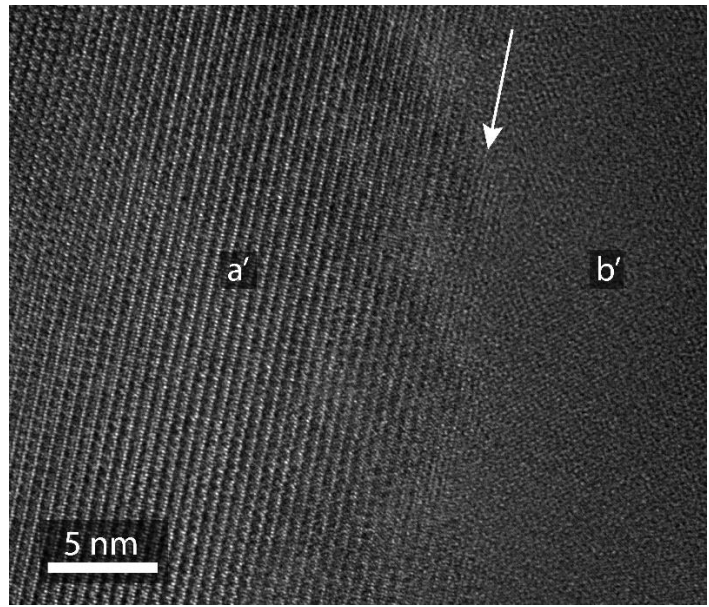

**Figure S1:** High-resolution transmission electron microscopy (TEM) image a grain boundary in $ErMnO_3$. The position of inclined grain boundary is indicated by the arrow, and the two adjacent grains (a' and b') are labelled and distinguishable. No intergranular phases can be observed at the grain boundary.

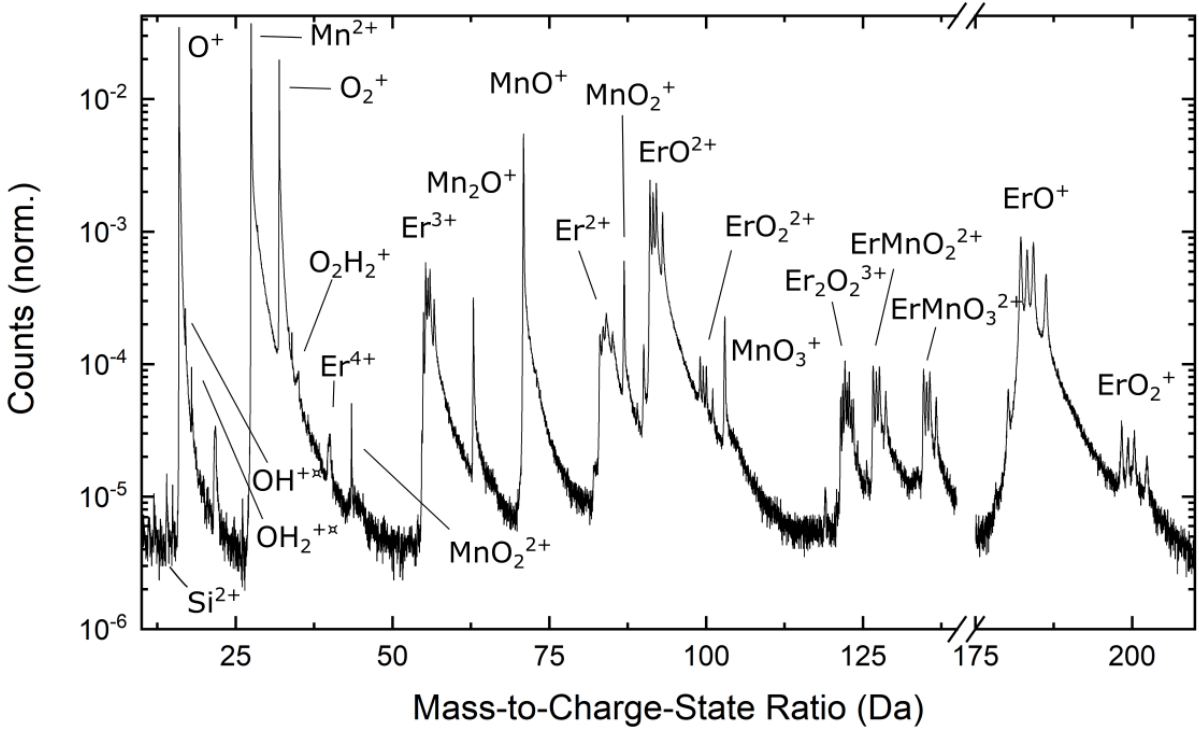

**Figure S2: Mass spectrum of ErMnO₃.** The figure shows the histogram of the mass-to-charge-state ratio obtained from the region of the grain boundary shown in Figure 3. The ionic species used in the reconstruction are labelled. ¤ indicates ionic species containing H which does not belong to the specimen and are not ranged.

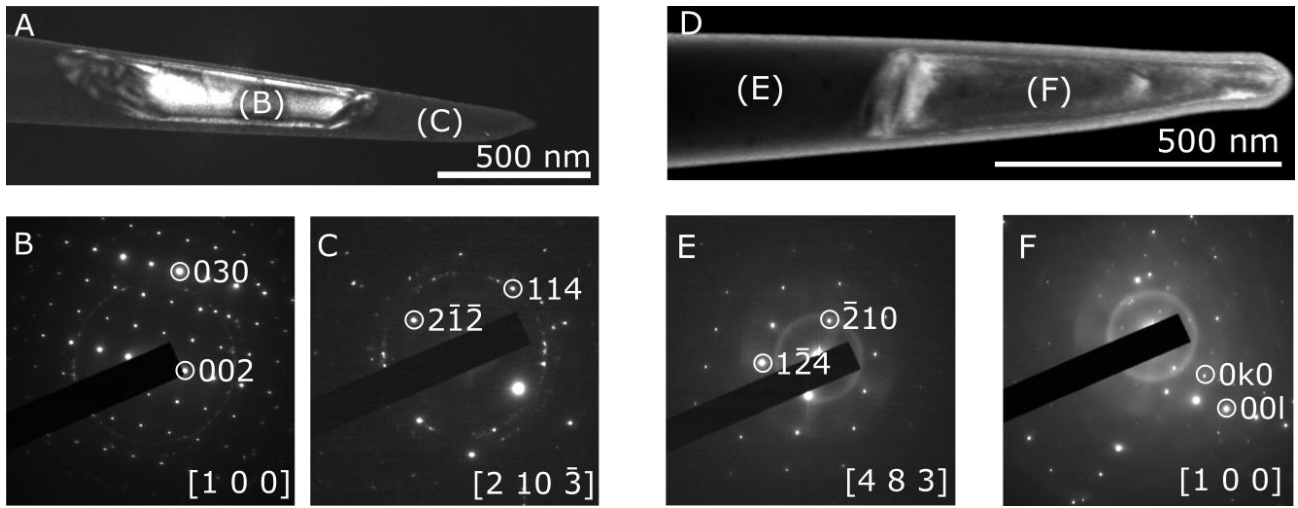

**Figure S3: TEM data from grain boundaries used for this study.** (A and D) shows dark-field (DF) images of the grain boundaries using a reflection from only one grain to create contrast between the grains: 002 for grain (B) in A and 00l of grain (F)in B). (B, C and E, F) shows respective diffraction patterns from the two grains in each specimen, where one of the grains is oriented along a major zone axis. A shaded box is placed behind the label (B) in subfigure (A) to enhance the contrast of the label. Note that (F) contains several reflections not consistent with the [100] zone axes, which are not considered to originate in the primary grain.

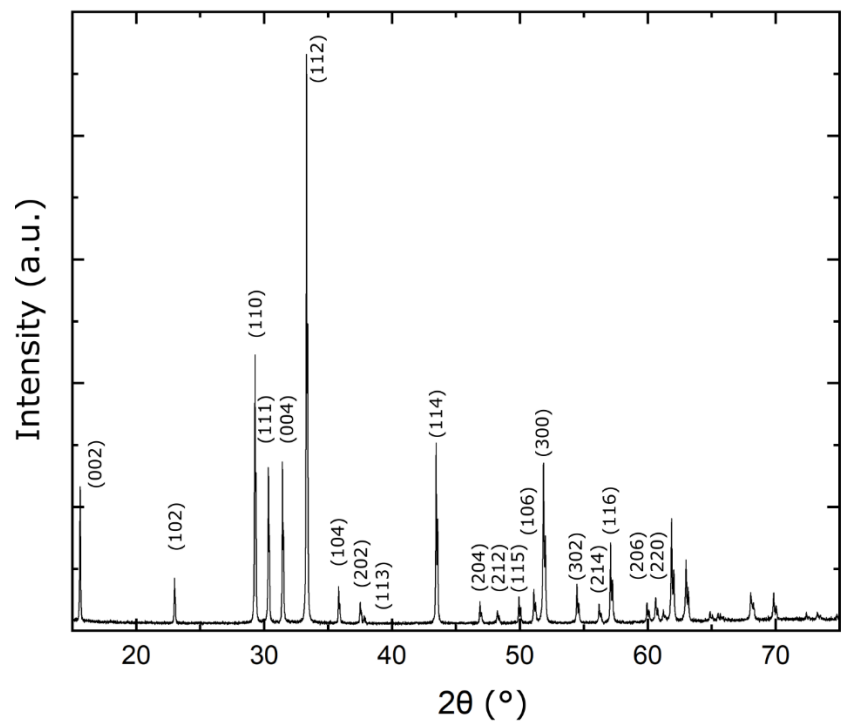

**Figure S4: X-ray diffraction characterization of polycrystalline ErMnO$_3$.** The experimental X-ray diffraction (XRD) pattern obtained on the polycrystalline samples. The main peaks are identified according to the hexagonal *P6$_3$cm* phase (see ref. 21).